\documentclass[aps,pra,reprint,showpacs]{revtex4-1}

\usepackage[english]{babel}
\usepackage[T1]     {fontenc}
\usepackage[utf8]   {inputenc}

\usepackage{amsmath}
\usepackage{amssymb}

\usepackage{bm}

\usepackage         {units}
\usepackage[squaren]{SIunits}

\usepackage{graphicx}

\usepackage{acronym}

\usepackage[version=3]{mhchem}

\usepackage{hyperref}

\newcommand{\abs}[1]{\ensuremath{\left\vert #1 \right\vert}}

\newcommand{\ee}{\ensuremath{\mathrm{e}}}
\newcommand{\eto}[1]{\ensuremath{\,{\ee}^{#1}}}
\newcommand{\ii}{\ensuremath{\mathrm{i}}}
\newcommand{\matrixel}[3]{\ensuremath{\left\langle \left. \vphantom{#2 #3}#1 \right\vert #2 \left\vert \vphantom{#2 #1} #3 \right.\right\rangle}}
\newcommand{\pt}{\ensuremath{\mathcal{PT}}}
\newcommand{\qmprod}[2]{\ensuremath{\left\langle #1\middle\vert #2 \right\rangle}}
\renewcommand{\vec}[1]{{\ensuremath{\bm{#1}}}}
\DeclareMathOperator{\diag}{diag}

\begin{document}

\title{Hermitian four-well potential as a realization of a
  \texorpdfstring{$\pt$}{PT}-symmetric system}
\author{Manuel Kreibich}
\email{Manuel.Kreibich@itp1.uni-stuttgart.de}
\author{J\"org Main}
\author{Holger Cartarius}
\author{G\"unter Wunner}
\date{\today}
\affiliation{1. Institut f\"ur Theoretische Physik, Universit\"at
  Stuttgart, Pfaffenwaldring 57, 70550 Stuttgart, Germany}

\begin{abstract}
  A $\pt$-symmetric Bose-Einstein condensate can be theoretically
  described using a complex optical potential, however, the
  experimental realization of such an optical potential describing the
  coherent in- and outcoupling of particles is a nontrivial task.  We
  propose an experiment for a quantum mechanical realization of a
  $\pt$-symmetric system, where the $\pt$-symmetric currents of a
  two-well system are implemented by coupling two additional wells to
  the system, which act as particle reservoirs. In terms of a simple
  four-mode model we derive conditions under which the two middle
  wells of the Hermitian four-well system behave \emph{exactly} as the
  two wells of the $\pt$-symmetric system. We apply these conditions
  to calculate stationary solutions and oscillatory dynamics.  By
  means of frozen Gaussian wave packets we relate the Gross-Pitaevskii
  equation to the four-mode model and give parameters required for the
  external potential, which provides approximate conditions for a
  realistic experimental setup.
\end{abstract}


\maketitle

% Abbreviations
\acrodef{BEC}{Bose-Einstein condensate}
\acrodef{GPE}{Gross-Pitaevskii equation}

In quantum mechanics an observable is described by an Hermitian
operator. This is true in particular for the energy, which is
represented by the Hamiltonian. The Hermicity is sufficient for purely
real eigenvalues, but is this really a necessary condition? Bender and
Boettcher found that for non-Hermitian Hamiltonians with a weaker
condition, namely $\pt$ symmetry, there exist parameters for which the
energy eigenvalue spectrum is purely real \cite{Bender98}, where $\pt$
stands for a combined action of parity $\mathcal{P}$ ($x \to -x$, $p
\to -p$), and time reversal $\mathcal{T}$ ($x \to x$, $p \to -p$ with
$\mathcal{T} \ii = - \ii \mathcal{T}$).

Due to the close analogy between the Schr\"odinger equation and the
equations describing the propagation of light in structured wave
guides, a $\pt$-symmetric optical system could be visualized
\cite{Klaiman08} and experimentally investigated
\cite{Guo09,Rueter10}. The necessary complex potential corresponds to
a complex refractive index, which is realized by balanced gain and
loss of light in the wave guide. Several other systems with $\pt$
symmetry have been suggested and partially realized, including lasers
\cite{Chong11,Ge11,Liertzer12}, electronics
\cite{Schindler11,Ramezani12,Schindler12}, microwave cavities
\cite{Bittner12}, and quantum field theories
\cite{Bender98,Bender12}. But up to date, a quantum mechanical
realization of a $\pt$-symmetric system is still missing.

It was proposed \cite{Klaiman08} that a system similar to complex
refractive indices in wave guides could be realized with \acp{BEC} in
double-well potentials, where particles are injected in one well and
removed from the other one. \acp{BEC} in $\pt$-symmetric double-well
potentials have been investigated in the Bose-Hubbard model and the
mean-field approximation
\cite{Graefe08b,Graefe08a,Graefe10,Cartarius12b,Cartarius12a}. In
these Refs., the $\pt$ symmetry is given by a complex potential which
fulfills the condition $V^*(x) = V(-x)$ and describes the coherent in-
and outcoupling of atoms into and from the system. It has already been
shown that a bidirectional coupling between two \acp{BEC} is possible
and at the same time particles may be continuously ejected
\cite{Shin05,Gati06}.

\begin{figure}
  \centering
  \includegraphics[width=0.47\columnwidth]{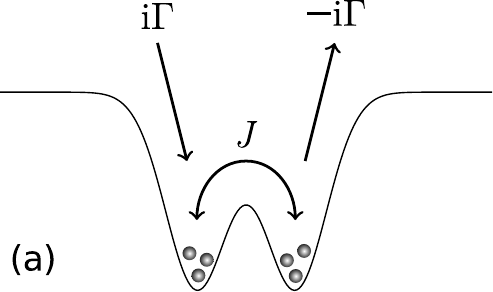} \hfill
  \includegraphics[width=0.47\columnwidth]{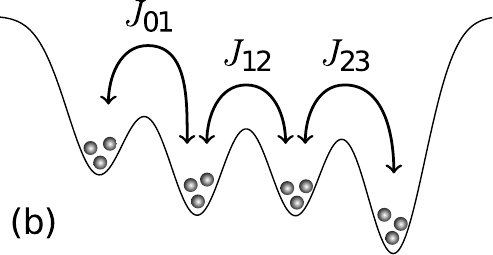}
  \caption{(a) $\pt$-symmetric two-well system with tunneling rate $J$
    and complex potential $\pm \ii \Gamma$, which describes coherent
    in- and outcoupling of atoms. (b) Alternatively, two additional
    wells serve as particle reservoirs and induce particle currents
    that -- under application of specific system parameters to be
    determined -- are equivalent to the $\pt$-symmetric currents of
    system (a).}
  \label{fig:graphics}
\end{figure}

In this letter we follow a different approach. Instead of injecting
and removing particles, we couple two additional wells to a
double-well system. The tunneling of the outer wells can be used to
add and remove particles from the inner wells (see
Fig.~\ref{fig:graphics}). We investigate, if just considering the two
middle wells, they behave like the two wells of the $\pt$-symmetric
two-well system. For the four-well potential we assume a combination
of four Gaussian beams,
\begin{align}
  \label{eq:extpot}
  V_\text{ext}(\vec{r}) = \sum\limits_{i=0}^3 V_i \exp \left[ -
    \frac{2x^2}{w_x^2} - \frac{2y^2}{w_y^2} - \frac{2
      (z-s_z^i)^2}{w_z^2} \right],
\end{align}
with $V_i < 0$ the depth and $s_z^i$ the displacement along the $z$
axis of each well, and $w_\alpha$ the width of a single well in each
direction. The dynamics of a \ac{BEC} is described by the \ac{GPE}
\begin{align}
  \label{eq:gpe}
  \ii \hbar \partial_t \psi(\vec{r},t) = \left[ - \frac{\hbar^2}{2 m}
    \Delta + V_\text{ext}(\vec{r}) + N g \abs{\psi(\vec{r},t)}^2
  \right] \psi(\vec{r},t),
\end{align}
where $g = 4 \pi \hbar^2 a_\text{s} / m$ is the strength of the
nonlinearity, with $a_\text{s}$ being the scattering length.

It was shown that the $\pt$-symmetric two-mode model shows the
features specific of $\pt$-symmetric systems
\cite{Graefe12}. Therefore we use the simple, but instructive two- and
four-mode models for our investigations. For the moment we neglect
particle interaction, which will be taken into account later. The
Hamiltonian of the linear, $\pt$-symmetric two-mode model is given by
\begin{align}
  \label{eq:twomode}
  H^{(2)}=
  \begin{pmatrix}
    \ii \Gamma & -J \\
    -J & -\ii \Gamma
  \end{pmatrix},
\end{align}
which models a double-well system of non-interacting bosons. The real
quantity $J>0$ designates the tunneling amplitude between the two
wells, and $\Gamma \in \mathbb{R}$ gives the strength of the imaginary
$\pt$-symmetric potential, which models gain and loss,
respectively. Obviously, this Hamiltonian commutes with the combined
action of parity and time reversal, $[\pt,H^{(2)}]=0$, where the
parity operator is given in this matrix representation by
\begin{align}
  \label{eq:7}
  \mathcal{P}=
  \begin{pmatrix}
    0 & 1 \\
    1 & 0
  \end{pmatrix},
\end{align}
and time reversal is simply the complex conjugation. The eigenvalues
and eigenvectors are easily found to be $E_\pm = \pm
\sqrt{J^2-\Gamma^2}$ and $\psi_\pm = \left( \ii \Gamma \pm \sqrt{J^2 -
    \Gamma^2}, -J \right)^T$ (unnormalized). For $\Gamma < J$ the
eigenvalues are purely real and the eigenvectors obey $\pt$
symmetry. In the opposite case, $\Gamma > J$, the $\pt$ symmetry is
broken and the eigenvalues become purely imaginary. Thus, the simple
two-mode model features the basic properties of a $\pt$-symmetric
Hamiltonian.

To gain deeper insight, and for an easier comparison with the later
defined four-mode model, we give the time derivatives of the
observables of the system. Two observables are the number of particles
in each well, which are given by $n_k = \psi_k^* \psi_k$, where $\psi
\in \mathbb{C}^2$ describes a quantum mechanical state. With the
particle current $j_{12} = \ii J \left( \psi_1 \psi_2^* - \psi_1^*
  \psi_2 \right)$ between the two wells the time-scaled ($t \to \hbar
t$) Schrödinger equation can be brought to the closed set of
differential equations for the observables
\begin{subequations}
  \label{eq:obs2}
  \begin{gather}
    \partial_t n_1 = -j_{12} + 2 \Gamma n_1, \quad\quad
    \partial_t n_2 = +j_{12} - 2 \Gamma n_2, \\
    \partial_t j_{12} = 2 J^2 (n_1 - n_2).
  \end{gather}
\end{subequations}
The imaginary potential induces particle currents from and to the
environment, $j_{e1} = 2 \Gamma n_1$ and $j_{2e} = 2 \Gamma n_2$, both
of which are proportional to the number of particles in the
corresponding well.

It is now our main purpose to investigate, whether the behavior of the
$\pt$-symmetric two-mode model \eqref{eq:twomode} can be described by
a Hermitian four-mode model, where two additional wells are coupled to
the system. The Hamiltonian of the four-mode model is given by
\begin{align}
  \label{eq:fourmode}
  H^{(4)}(t)=
  \begin{pmatrix}
    E_0(t) & -J_{01}(t) & 0 & 0 \\
    -J_{01}(t) & 0 & -J_{12} & 0 \\
    0 & -J_{12} & 0 & -J_{23}(t) \\
    0 & 0 & -J_{23}(t) & E_3(t)
  \end{pmatrix}.
\end{align}
The two middle wells will be symmetric, hence $E_1=E_2=0$. The
tunneling amplitudes and on-site energies of the outer wells are
$J_{01}$, $J_{23}$, and $E_0$, $E_3$, respectively. They may be
time-dependent, as denoted by the explicit time-dependence in
Eq.~\eqref{eq:fourmode}. To be able to compare the four-mode model
with the two-mode model, we calculate the time derivatives of the
particle populations, which yields the simple relations
\begin{align}
  \label{eq:ndot}
  \partial_t n_0 &= - j_{01}, &
  \partial_t n_1 &= j_{01} - j_{12}, \nonumber \\
  \partial_t n_2 &= j_{12} - j_{23}, &
  \partial_t n_3 &= j_{23},
\end{align}
where
\begin{align}
  \label{eq:jdef}
  j_{kl} = \ii J_{kl} \left( \psi_k \psi_l^* - \psi_k^* \psi_l \right)
\end{align}
is the particle current between adjacent wells. By additionally
considering the time derivative of $j_{12}$, we obtain
\begin{subequations}
  \label{eq:obs4}
  \begin{gather}
    \partial_t n_1 = j_{01} - j_{12}, \quad\quad
    \partial_t n_2 = j_{12} - j_{23}, \\
    \partial_t j_{12} = 2 J_{12}^2 \left( n_2 - n_1 \right) + J_{12}
    \left( J_{23} C_{13} - J_{01} C_{02} \right),
  \end{gather}
\end{subequations}
where we defined $C_{kl} = \psi_k \psi_l^* + \psi_k^*
\psi_l$. Comparing Eqs.~\eqref{eq:obs2} and~\eqref{eq:obs4} we can
conclude that if the conditions
\begin{align}
  \label{eq:cond}
  j_{01} &= 2 \Gamma n_1, &
  j_{23} &= 2 \Gamma n_2, &
  J_{01} C_{02} &= J_{23} C_{13}
\end{align}
are fulfilled, the two middle wells of the Hermitian four-mode model
\eqref{eq:fourmode} behave \emph{exactly} as the two wells of the
$\pt$-symmetric two-mode system \eqref{eq:twomode}.

We now need to give the explicit time-dependency of the free
parameters of the Hamiltonian~\eqref{eq:fourmode}, $E_0$, $E_3$,
$J_{01}$ and $J_{23}$, such that Eqs.~\eqref{eq:cond} are
fulfilled. For the tunneling elements we can set
\begin{align}
  \label{eq:tunneldef}
  J_{01} = d C_{13}, && J_{23} = d C_{02},
\end{align}
where $d$ is a time-independent parameter, and can be tuned to bring
the tunneling elements into an experimentally realizable range. The
currents $j_{01}$ and $j_{23}$ have to fulfill
Eqs.~\eqref{eq:cond}. However, with the tunneling elements determined
by Eqs.~\eqref{eq:tunneldef}, there are no free parameters left to
adjust these currents. Instead, we take the time derivatives of
Eqs.~\eqref{eq:cond}, which yields
\begin{subequations}
  \label{eq:8}
  \begin{align}
    \partial_t j_{01} &= 2 \dot{\Gamma} n_1 + 2 \Gamma \dot{n}_1 = 2
    \dot{\Gamma} n_1 + 2 \Gamma \left( 2 \Gamma n_1 - j_{12} \right), \\
    \partial_t j_{23} &= 2 \dot{\Gamma} n_2 + 2 \Gamma \dot{n}_2 = 2
    \dot{\Gamma} n_2 + 2 \Gamma \left( j_{12} - 2 \Gamma n_2 \right).
  \end{align}
\end{subequations}
where we allow the parameter $\Gamma$ to be explicitly time-dependent.
For the last equalities we used Eqs.~\eqref{eq:ndot}
and~\eqref{eq:cond}. Here, $\Gamma$ is not a quantity entering the
Hamiltonian directly as in the two-mode model \eqref{eq:twomode}, but
a free parameter, which determines the matrix elements of the
Hamiltonian \eqref{eq:fourmode}.

Now we can calculate the time derivatives of $j_{01}$ and $j_{23}$
from the definition \eqref{eq:jdef}. This leads to the linear system
of equations for the on-site energies $E_0$ and $E_3$,
\begin{align}
  \label{eq:edef}
  \begin{pmatrix}
    J_{01} C_{01} & d \tilde{j}_{01} \tilde{j}_{13} \\
    -d \tilde{j}_{02} \tilde{j}_{23} & -J_{23} C_{23}
  \end{pmatrix}
  \begin{pmatrix}
    E_0 \\ E_3
  \end{pmatrix}
  =
  \begin{pmatrix}
    v_0 \\ v_3
  \end{pmatrix},
\end{align}
with the entries
\begin{subequations}
  \begin{gather}
    \label{eq:11}
    v_0 = 2 \dot{\Gamma} n_1 + 2\Gamma \left( j_{01} - j_{12}
    \right) - 2 J_{01}^2 \left( n_0 - n_1 \right) \nonumber \\
    - J_{01} J_{12} C_{02} - d \left( J_{01} \tilde{j}_{03} + J_{12}
      \tilde{j}_{23}- J_{23} \tilde{j}_{12} \right) \tilde{j}_{01},
    \displaybreak[0] \\
    v_3 = 2 \dot{\Gamma} n_2 + 2 \Gamma \left( j_{12} - j_{23}
    \right) - 2 J_{23}^2 \left( n_2 - n_3 \right) \nonumber \\
    + J_{23} J_{12} C_{13} - d \left( J_{01} \tilde{j}_{12} - J_{12}
      \tilde{j}_{01} - J_{23} \tilde{j}_{03} \right) \tilde{j}_{23}.
  \end{gather}
\end{subequations}
Here we have defined the modified currents $\tilde{j}_{kl} = \ii
\left( \psi_k \psi_l^* - \psi_k^* \psi_l \right)$. Thus, the on-site
energies $E_0$ and $E_3$ are used to maintain the validity of
Eqs.~\eqref{eq:cond}. Since the energies do not determine the currents
but their time derivatives, the initial wave function must be chosen
in such a way that the conditions are fulfilled.

So far we could find the explicit time-dependences of the matrix
elements $E_0(t)$, $E_3(t)$, $J_{01}(t)$, and $J_{23}(t)$ of the
four-mode model in order that at every time the two middle wells have
the same behavior as the $\pt$-symmetric two-mode model. Thus our
method is a valid possibility to realize a $\pt$-symmetric quantum
mechanical system. To calculate the matrix elements at every time step
we integrate the four-dimensional complex Schrödinger equation with a
numerical integrator and use Eqs.~\eqref{eq:tunneldef}
and~\eqref{eq:edef}. We now give two examples for different solutions.

\begin{figure}[tb]
  \centering
  \includegraphics[width=\columnwidth]{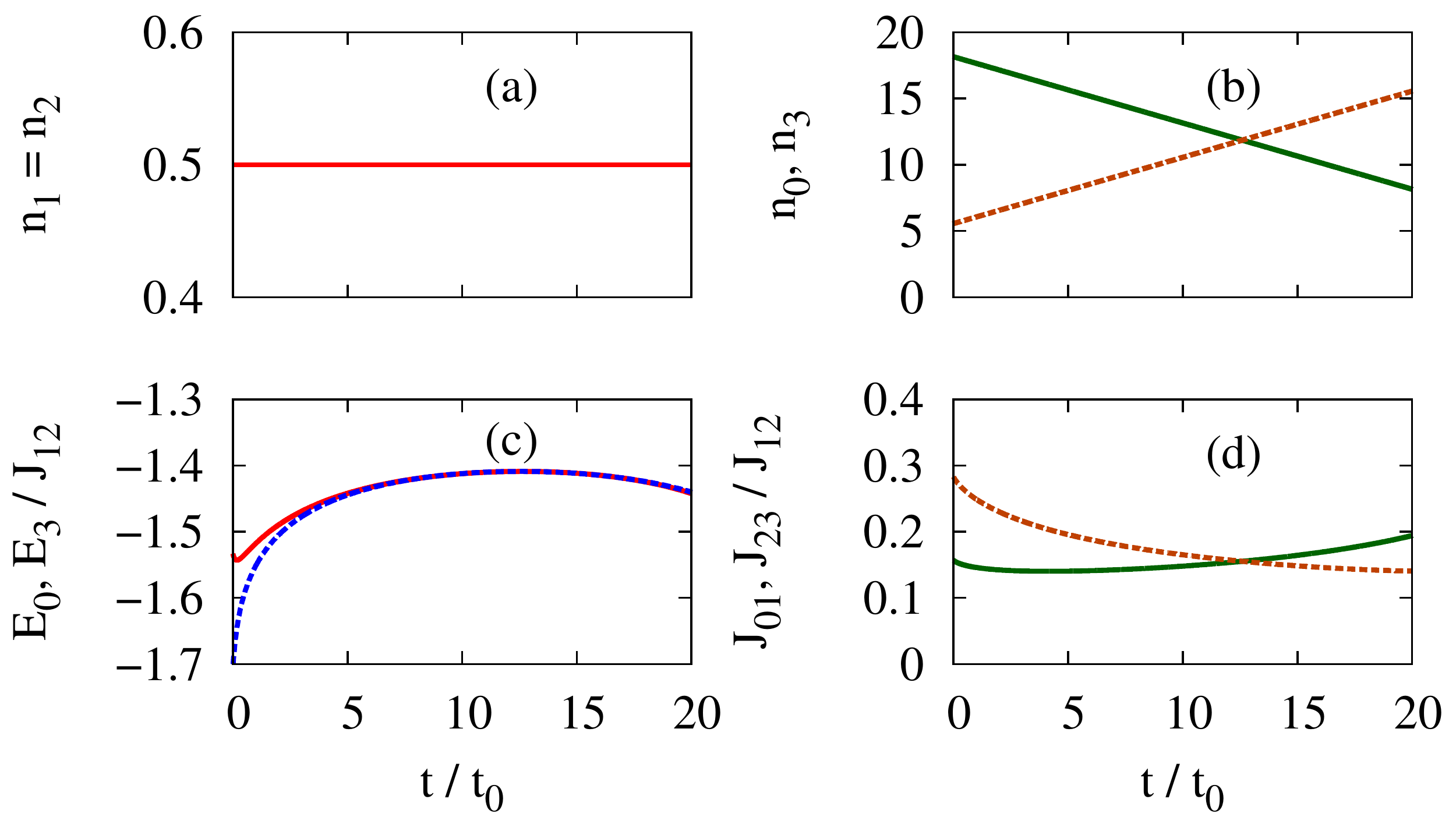}
  \caption{(Color online) (a) Populations $n_1 = n_2 = 1/2$ of the
    middle wells, which are both equal and constant in time. (b)
    Populations $n_0$ (solid) and $n_3$ (dashed) of the outer wells,
    which depend linearly on the time, i.\,e.\ $\dot{n}_0 = -\Gamma$
    and $\dot{n}_3 = \Gamma$. (c) Time-dependent on-site energies
    $E_0$ (solid) and $E_3$ (dashed). (d) Tunneling amplitudes
    $J_{01}$ (solid) and $J_{23}$ (dashed). Time is given in units of
    $t_0 = \hbar / J_{12}$.}
  \label{fig:stat_sol}
\end{figure}

First we consider a quasi-stationary solution, i.\,e.\ a state in
which the particle numbers in the two middle wells are
stationary. These states correspond to the stationary solution of the
two-mode model. We prepared this stationary solution for $\Gamma /
J_{12} = 0.5$ at $t=0$. Fig.~\ref{fig:stat_sol} shows the results. As
required, the number of particles in the middle wells are equal and
constant in time (we have chosen the normalization such that
$n_1+n_2=1$). From Eqs.~\eqref{eq:ndot} and~\eqref{eq:cond} we then
obtain $\dot{n}_0 = - \Gamma$ and $\dot{n}_3 = \Gamma$, plotted in
Fig.~\ref{fig:stat_sol}(b). Figs.~\ref{fig:stat_sol}(c) and~(d) show
the calculated matrix elements. All of them vary only slightly in
time. Due to the linear decrease of the number of particles in well 3
the available time for the $\pt$ symmetry is limited.

\begin{figure}[tb]
  \centering
  \includegraphics[width=\columnwidth]{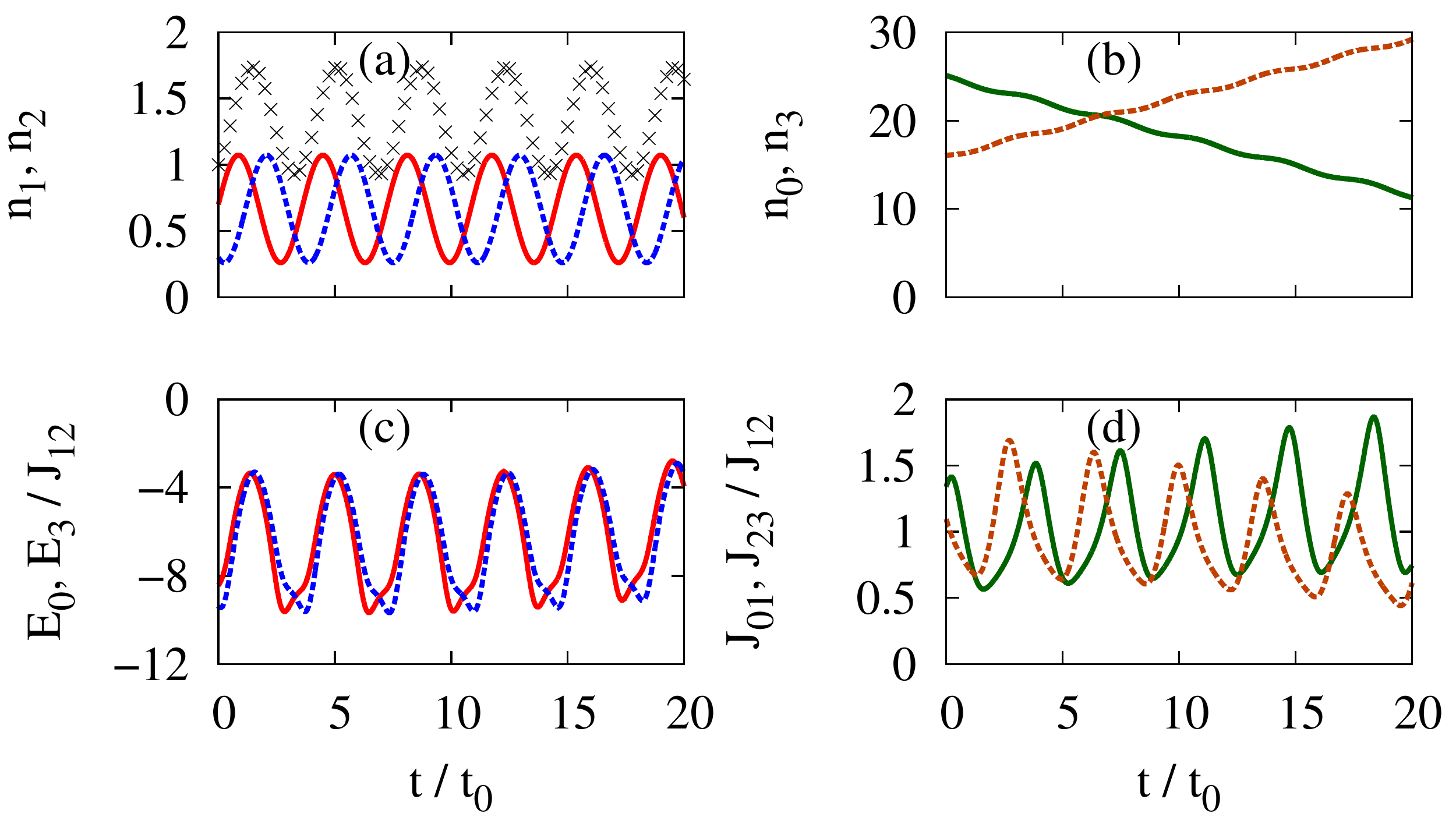}
  \caption{(Color online) Same as Fig.~\ref{fig:stat_sol}, but for
    oscillatory dynamics. Additionally, the total number of particles
    in wells~1 and~2 are plotted as black crosses in (a).}
  \label{fig:dynamics}
\end{figure}

As a second example we prepared a non-stationary solution at $t=0$ for
$\Gamma / J_{12} = 0.5$ (Fig.~\ref{fig:dynamics}). There are the
typical Rabi-type oscillations, but with a smaller phase difference
$\Delta \phi < \pi$, which leads to a non-constant added number of
particles in the two middle wells. This value oscillates harmonically
around its mean value. The same behavior is obtained for the optical
system in Ref.~\cite{Klaiman08}. The matrix elements show a
quasi-oscillatory behavior. As before, the time for an exact $\pt$
symmetry is limited.

\begin{figure}
  \centering
  \includegraphics[width=0.8\columnwidth]{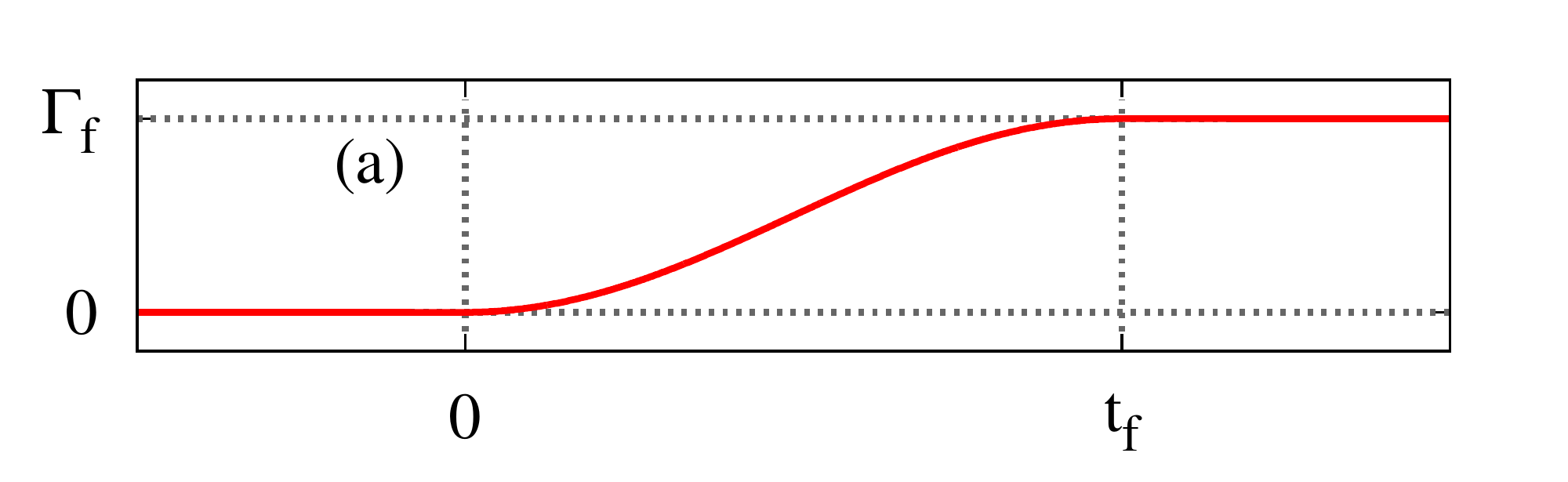}
  \includegraphics[width=\columnwidth]{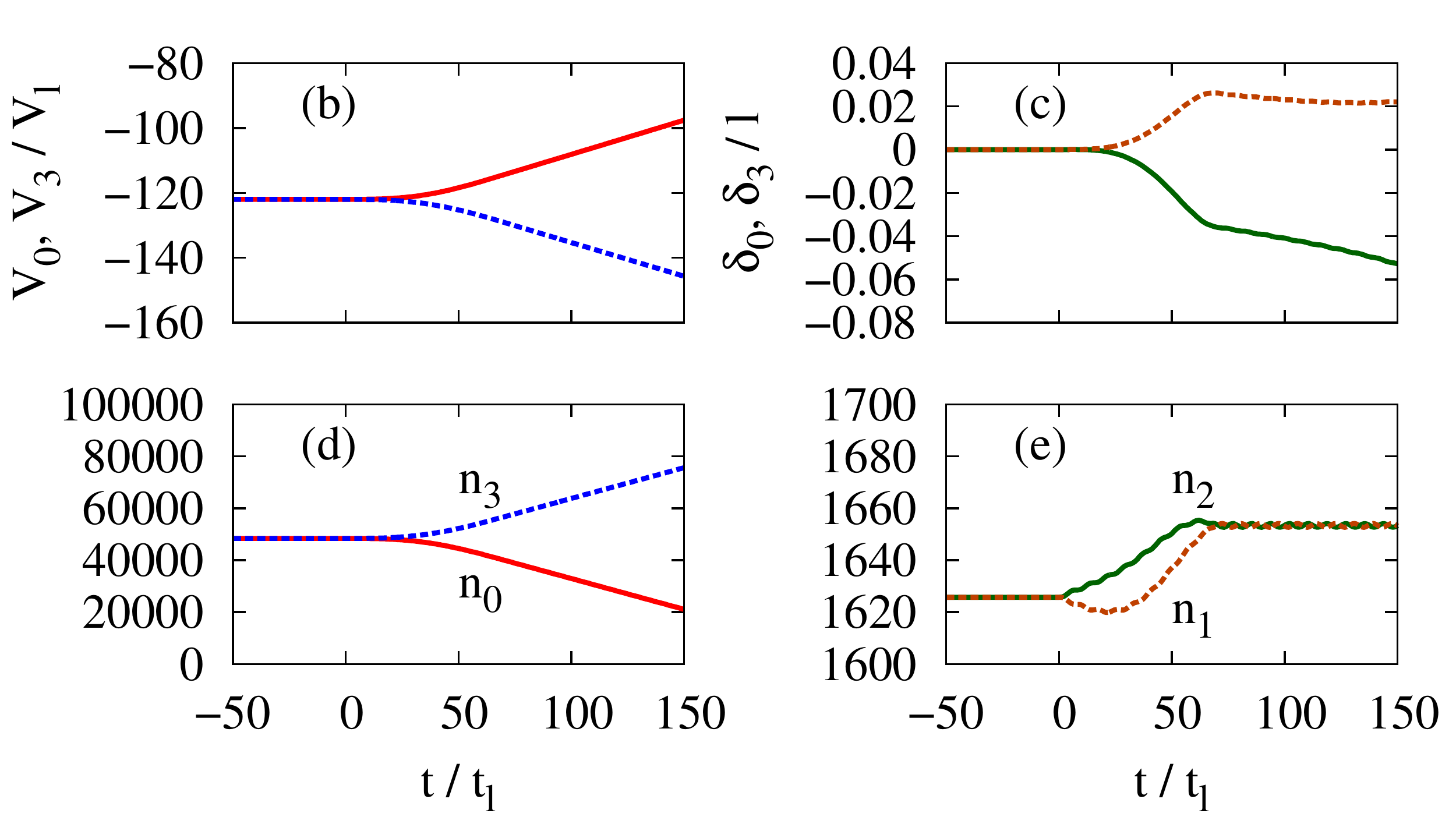}
  \caption{(Color online) (a) Adiabatic ramp of the current from
    $\Gamma=0$ to a target value of $\Gamma_\text{f}$ as a function of
    the time $t$. Time dependences of (b) the potential depths $V_0$
    (solid) and $V_3$ (dashed) and (c) the displacements of the outer
    wells $\delta_0$ (solid) and $\delta_3$ (dashed). (d) and (e)
    Number of particles in each well.}
  \label{fig:adiab}
\end{figure}

For these calculations, the initial conditions had to be chosen
appropriately so that they obey $\pt$ symmetry. This would be a
difficult task in an experiment. For that, we propose an approach of
adiabatically increasing the $\pt$ parameter $\Gamma$. We start at the
ground state of the four-well system, and increase $\Gamma$ according
to $\Gamma(t) = \Gamma_\text{f} \left[ 1 - \cos\left( \pi t /
    t_\text{f} \right) \right] / 2$ for $t \in [0,t_\text{f}]$ (see
Fig.~\ref{fig:adiab}a)). The quantities $\Gamma_\text{f}$ and
$t_\text{f}$ have to be chosen such that $\dot{\Gamma} / \Gamma \ll
\omega$, where $\omega$ is the typical oscillation frequency. The
system then changes adiabatically from the ground state of the
Hermitian system to a $\pt$-symmetric ground state.

So far we have neglected the contact interaction of the
atoms. However, as we show next, taking into account the nonlinear
contact interaction will not change the characteristic $\pt$-symmetric
behavior. The Hamiltonian, which is nonlinear in the mean-field
approximation, then reads
\begin{align}
  \label{eq:2}
  H^{(4)}(t) = H^{(4)}_\text{lin}(t) + c \diag \left( \abs{\psi_1}^2,
    \abs{\psi_2}^2, \abs{\psi_3}^2, \abs{\psi_4}^2 \right),
\end{align}
where $H^{(4)}_\text{lin}$ is the linear part \eqref{eq:fourmode}. The
quantity $c$ measures the strength of the interaction. We have to
recalculate the time derivatives of the observables. For the
derivatives of the populations, $\dot{n}_1$ and $\dot{n}_2$, we obtain
the same as in the linear case. For the current between the middle
wells we obtain
\begin{subequations}
  \begin{align}
    \label{eq:12}
    &\text{Two wells:} & \partial_t j &= \left( \partial_t j
    \right)_\text{lin} + J c \left( n_1 - n_2 \right) C_{12}, \\
    &\text{Four wells:} & \partial_t j_{12} &= \left( \partial_t
      j_{12} \right)_\text{lin} + J_{12} c \left( n_1 - n_2 \right)
    C_{12}.
  \end{align}
\end{subequations}
Generally, the time evolution of $C_{12}$ differs for the two- and
four-mode model. But in the case of adiabatically increasing $\Gamma$,
we have $n_1 \approx n_2$, i.\,e., with the choice of parameters given
above we have an \emph{approximate} equivalence of the two- and
four-mode model. Solely the linear system of equations \eqref{eq:edef}
has to be modified to include the interaction.

We further want to give approximate parameters for a realistic
potential. Four wells can be realized by a superposition of four
Gaussian laser beams \eqref{eq:extpot}. The dynamics of a \ac{BEC} is
described by the \ac{GPE} \eqref{eq:gpe}. To relate the four-mode
model to the \ac{GPE}, we assume the wave function to be a
superposition of frozen Gaussian wave packets,
\begin{align}
  \label{eq:varansatz}
  \psi = \sum\limits_{k=0}^3 d^k g^k = \sum\limits_{k=0}^3 d^k
  \eto{-A_x x^2 - A_y y^2 - A_z (z-q_z^k)^2}.
\end{align}
For simplicity we assume the widths $A_\alpha$ to be constant in time,
and the same for each Gaussian, and the displacement of each Gaussian
to be the same as the displacement of the corresponding well, $q_z^i =
s_z^i$. Only the amplitudes of the Gaussians $d^k$, are dynamical
variables.

Multiplying Eq.~\eqref{eq:gpe} by $\psi^*$ and integrating over
$\mathbb{R}^3$, we obtain the equations of motion for $d^k$,
\begin{align}
  \label{eq:3}
  \ii \hbar \sum\limits_{k=0}^3 \qmprod{g^l}{g^k} \dot{d}^k =
  \sum\limits_{k=0}^3 \matrixel{g^l}{\hat{H}}{g^k} d^k.
\end{align}
With the method of symmetric orthogonalization \cite{Lowdin50} we can
write these equations as a Schrödinger equation with a symmetric and
Hermitian $4 \times 4 $ Hamiltonian. By considering only nearest
neighbors in the integrals, we can relate the matrix elements of the
four-mode model \eqref{eq:fourmode} to the realistic potential
\eqref{eq:extpot}. This yields
\begin{subequations}
  \begin{align}
    \label{eq:4}
    E_k &= \frac{\hbar^2}{2 m} \left( A_x + A_y + A_z \right) + V_k
    \beta_x \beta_y \beta_z, \displaybreak[0] \\
      J_{lk} &= \frac{\hbar^2}{2 m} A_z^2 \left( q_z^l - q_z^k
      \right)^2 \gamma + \left( V_l + V_k \right) \beta_x \beta_y
      \beta_z \gamma \nonumber \\
      &\hphantom{= \frac{\hbar^2}{2 m} A_z^2 \left( q_z^l - q_z^k
        \right)^2} \times \left( \frac{1}{2} - \gamma^{1/(1+A_z
          w_z^2)} \right), \displaybreak[0] \\
    c &= \frac{4 \hbar^2 N a_\text{s}}{m} \sqrt{\frac{A_x A_y
        A_z}{\pi}},
  \end{align}
\end{subequations}
with the abbreviations
\begin{align}
  \label{eq:5}
  \beta_\alpha &= \sqrt{\frac{A_\alpha w_\alpha^2}{1 + A_\alpha
      w_\alpha^2}}, &
  \gamma &= \exp \left[ - \frac{A_z}{2} \left( q_z^l - q_z^k
    \right)^2 \right].
\end{align}
The parameters $A_\alpha$ are determined by minimizing the mean-field
energy.

We calculated the adiabatic current ramp for a condensate of $N=10^5$
atoms of \ce{^{87}Rb} with a scattering length tuned to $a = 10.9
a_\text{B}$ ($a_\text{B}$ being the Bohr radius). The distance between
the middle wells is $l=\unit{2}{\micro\meter}$. We express the
distance of the outer wells by their deviation from a equidistant
lattice, i.\,e.\ $q_z^0 = -3l/2 + \delta^0$ and $q_z^3 = 3l/2 +
\delta^3$. The basic unit of energy is $E_l = \hbar^2 / m l^2$, which
yields $E_l / h = \unit{29.1}{\hertz}$. As initial condition we use
the ground state for $V_1 = V_2 = -80 E_l$ and $V_0 = V_3 = -122
E_l$. The parameter $d$ in Eq.~\eqref{eq:tunneldef} is chosen such
that $\delta^0 = \delta^3 = 0$ for $t \leq 0$. We have chosen the
widths of the trap \eqref{eq:extpot} to be $w_x=w_y=4l$ and
$w_z=l/2$. The basic unit of time is $t_l = \hbar / E_l =
\unit{5.47}{\milli\second}$. Fig.~\ref{fig:adiab} shows the results
for $\Gamma_f / J_{12} = 0.5$ and $t_f / t_l \approx 70$.

The trap depth $V_0^0$ has to be increased, whereas $V_0^3$ has to be
decreased, in an almost symmetric way. The distance of the outer wells
has to be varied within a few percent of the distance $l$. The results
of the particle numbers in each well confirm that the change of
$\Gamma$ is approximately adiabatic (see Fig.~\ref{fig:adiab}). For
these conditions, the system arrives at an approximate $\pt$-symmetric
ground state for $t \gtrsim 70 t_l$, which can be determined by the
constant number of particles in wells $1$ and $2$. We note that the
change of the distances of the outer wells can be neglected, as
further calculations show. Furthermore, the simulation indicates that
the system is robust with respect to small random perturbations of the
external potential.

To summarize we have shown that the two middle wells of the Hermitian
four-mode model can show the same behavior as the $\pt$-symmetric
two-mode model, and thus offers an approach to realize $\pt$ symmetry
in a quantum mechanical system. This agreement is \emph{exact} in the
absence of interaction. We have proposed the method of adiabatically
increasing the $\pt$ parameter $\Gamma$ to create the $\pt$-symmetric
ground state. This works also \emph{approximately} for particles with
interaction in the mean-field limit. We finally estimated parameters
of a realistic potential, which would be necessary to prepare such an
experiment. The time-dependent potential \eqref{eq:extpot} could be
realized e.\,g.\ using an acousto-optical modulator
\cite{Henderson09}. For future work it is desirable to extend these
investigations to a \ac{BEC} described by the full \ac{GPE} to obtain
more accurate parameters for an experimental realization.

\begin{acknowledgments}
  This work was supported by DFG\@. M.\,K. is grateful for support
  from the Landesgraduiertenförderung of the Land Baden-Württemberg.
\end{acknowledgments}

\end{document}